\documentstyle[aps,pra,twocolumn,epsfig]{revtex}
\begin{document}
\draft
\title{Absorption by cold Fermi atoms in a harmonic trap}
\author{Gediminas Juzeli\={u}nas and Marius 
Ma\v{s}alas} 
\address{Institute of Theoretical Physics and Astronomy, A. Go\v{s}tauto 12, \\
Vilnius 2600, Lithuania }
\date{\today}
\maketitle
\begin{abstract}
We study the absorption spectrum for a strongly degenerate Fermi gas
confined in a harmonic trap. The spectrum is calculated using 
both the exact summation and also the Thomas-Fermi (TF) approximation.
In the latter case, relatively simple analytical expressions are
obtained for the absorption lineshape at
large number of trapped atoms. At zero
temperature, the approximated lineshape 
is characterized by a $\left(1-z^2\right)^{5/2}$ dependence which agrees well 
with the exact numerical calculations. At
non-zero temperature, the spectrum becomes broader, although remains
non-Gaussian as long as the fermion gas is degenerate. The changes in the 
trap frequency for an electronically excited atom can introduce 
an additional line broadening.
\end{abstract}

In recent years there has been a great deal of interest in the dilute gas of
trapped atoms cooled to temperatures below $1$ $\mu {\rm K}$. At such low
temperatures, an important role is played by the quantum statistics of
atoms. Bosons tend to occupy the lowest translational level of the trap to
form the Bose - Einstein condensate \cite{anders:95,Ketterle,Hulet}. On the
other hand, fermions can still occupy many trap levels (predominantly up to
the Fermi level), as it was observed recently \cite{demarco:99}. The physical
properties of a Fermi gas (such as the specific heat) depend on the number of
atoms in the system \cite{Butts:97,Wallis}, as well as on the trap
anisotropy \cite{Wallis}.

The optical spectroscopy has proven to be useful in getting information
about cold atomic gases \cite
{anders:95,Ketterle,Hulet,demarco:98,Lewenst:94,Lewenst:99,javan}. The
Fermi-Dirac (FD) statistics is known to change the optical response of the
system compared to the classical one. The signatures of quantum degeneracy
emerge in the scattering spectra of homogeneous\ \cite{javan} and trapped 
\cite{demarco:98} Fermi gases. Effects of quantum statistics are also 
featured in
the scattering of short laser pulses from a trapped Fermi gas \cite
{Lewenst:99}. Furthermore, the spontaneous emission appears to be inhibited
in a cold Fermi gas \cite{Busch}.

The effects of quantum degeneracy should manifest in the absorption spectra
as well. The aim of the present paper is to investigate absorption by a cold
Fermi gas confined in a harmonic trap. The analysis concentrates on the
degenerate Fermi gas (i.e. very low temperatures), for which the quantum
statistics of the atoms plays an important role. The theory involves exact
calculations, as well as the Thomas - Fermi (TF) approximation.
Consequently, the absorption spectrum is analyzed both for small and large
numbers of trapped atoms.

Consider a gas of Fermi atoms confined in a harmonic trap. The harmonic
approximation is relevant for the traps used in recent experiments \cite
{demarco:99}. We shall neglect atomic collisions, since the s-wave
collisions are forbidden between the spin - polarized fermions. 
Consequently, one can make use of the following one - atom Hamiltonian: 
\begin{equation}
H_{1-at}=\left| g\right\rangle H_{g}\left\langle g\right| +\sum_{ex}\left|
ex\right\rangle \left( \hbar \omega _{0,ex}+H_{ex}\right) \left\langle
ex\right| ,  \label{eq:1}
\end{equation}
where $\left| g\right\rangle $ and $\left| ex\right\rangle $ represent the
ground and an excited electronic state of an atom, $\hbar \omega _{0,ex}$%
{\em \ } is the excitation energy. It is noteworthy that the ground
electronic level of the fermion atom has a number of magnetic sublevels over
which the summation is to be carried out in the Hamiltonian (\ref{eq:1}).
However, such a summation is not necessary if the atoms are spin - polarized, as
it is the case in the experiment by DeMarco and Jin on $^{40}$K atoms 
\cite{demarco:99}. Here
also $H_{g}$ $\left( H_{ex}\right) $ is the Hamiltonian for the
translational motion of a trapped atom in the ground (excited) electronic{\em \ }%
state: 
\begin{equation}
H_{g,ex}\left({\bf r},{\bf p}\right)=\frac{p^{2}}{2M}+\frac{M\Omega _{g,ex}^{2}\left(
x^{2}+\lambda _{y}^{2}y^{2}+\lambda _{z}^{2}z^{2}\right) }{2},
\label{eq:2}
\end{equation}
where ${\bf p}=-i\hbar {\bf \nabla }$ 
is the momentum operator,  
$M$ is the atomic mass, 
$\Omega _{g}$ $\left( \Omega _{ex}\right) $ is the frequency
of the translational motion along the x axis if the atom is 
in the ground (excited)
electronic state, and the dimensionless parameters $\lambda _{y}$ and $%
\lambda _{z}$ describe the extent of anisotropy of the trap. Note that the
frequency $\Omega _{g}$\ can be generally different from $\Omega _{ex}$ 
due to the changes in the magnetic moment of the atom following 
its transition to an excited electronic state. The
effects related to this fact will be explored using the TF
approximation. 

The lineshape of the absorption spectrum is given by 
\begin{equation}
I\left( \omega \right) =\sum_{i,f}\rho _{i}\left| \left\langle f\right|
V\left| i\right\rangle \right| ^{2}\delta \left( \omega -\omega _{fi}\right)
.  \label{eq:3}
\end{equation}
Here $\left| i\right\rangle \equiv \left| g\right\rangle \left| {\bf n}%
\right\rangle _{g}$ and $\left| f\right\rangle \equiv \left| ex\right\rangle
\left| {\bf m}\right\rangle _{ex}$ are the initial and final states of an
atom, $\left| {\bf n}\right\rangle _{g}\equiv \left|
n_{x},n_{y},n_{z}\right\rangle _{g}$ and $\left| {\bf m}\right\rangle
_{ex}\equiv \left| m_{x},m_{y},m_{z}\right\rangle _{ex}$ are the atomic
translational states characterized by the energies 
\begin{equation}
\varepsilon _{{\bf n}}^{g,ex}=\hbar \Omega _{g,ex}\left( n_{x}+\lambda
_{y}n_{y}+\lambda _{z}n_{z}\right) +\varepsilon _{{\bf 0}}^{g,ex},
\end{equation}
where $\varepsilon _{{\bf 0}}^{g,ex}=\hbar \Omega _{g,ex}\left( 1+\lambda
_{y}+\lambda _{z}\right) /2$, and $\omega _{fi}=\omega _{0ex}+\left( \varepsilon
_{{\bf m}}^{ex}-\varepsilon _{{\bf n}}^{g}\right) /\hbar $ is the transition
frequency. Here also 
$\rho _{i}\equiv \rho \left( \varepsilon _{{\bf n}}^{g}\right) =\left[ \exp%
\left( \beta \varepsilon _{{\bf n}}^{g}-\beta \mu \right) +1\right] ^{-1}$
is the FD distribution function for the trapped atoms, $\mu $
is the chemical potential and $\beta =1/k_{B}T$. The operator 
\begin{equation}
V=\sum_{ex}\left\{ \left| ex\right\rangle d_{ex}\exp \left( i{\kappa x}%
\right) \left\langle g\right| +h.c.\right\}   \label{eq:5}
\end{equation}
describes the interaction between an atom and the electromagnetic field
propagating along the x axis, ${\kappa }$ is the wave number of the light, and $%
d_{ex}$ is the atomic transition dipole moment along the polarization of
the light.

Consider first the absorption spectrum using the exact summation over the 
translational levels. At this
stage, it is assumed that $\Omega _{g}=\Omega _{ex}=\Omega $, 
yet the trap can
still be anisotropic. At zero temperature, only the
levels with $n_{x}+\lambda _{y}n_{y}+\lambda _{z}n_{z}\leq n_{F}=E_{F}/\hbar
\Omega $ are occupied by the atoms, where $E_{F}\equiv \left. \mu \right|
_{T=0}$ is the Fermi energy. In such a situation, the absorption lineshape
takes the form: 
\begin{eqnarray}
I\left( \omega \right)  &=&\sum_{ex}\left| d_{ex}\right|
^{2}\sum_{n_{x}=0}^{n_{F}}\sum_{m=-n_{x}}^{\infty }K_{x}\delta \left( \omega
-\omega _{0,ex}-m\Omega \right)
\nonumber \\
&&\times n_{x}!\left( n_{x}+m\right) !e^{-\alpha ^{2}}
\left( \alpha ^{2}\right) ^{2n_{x}+m}
\nonumber \\
&&\times 
\left( \sum_{j=0}^{j_x }
\frac{\left( -\alpha ^{-2}\right)^{j}}{%
j!\left( n_{x}-j\right) !\left( n_{x}+m-j\right) !}\right) ^{2}
\label{I-tiksl},
\end{eqnarray}
with $j_x =\min \left(n_{x},n_{x}+m\right)$, where
$\alpha =\kappa \left( \hbar /2M\Omega \right) ^{1/2}$, and the
factor $K_{x}=\sum_{n_{y}=0}^{\left[ \frac{1}{\lambda _{y}}\left(
n_{F}-n_{x}\right) \right] }\sum_{n_{z}=0}^{\left[ \frac{1}{\lambda _{z}}%
\left( n_{F}-n_{x}-\lambda _{y}n_{y}\right) \right] }1$ represents a
number of occupied translational states $\left|
n_{x},n_{y},n_{z}\right\rangle $ for a fixed value of $n_{x}$, the brackets $%
\lbrack ...\rbrack $ labeling the integer part of a number.
The Fermi number $n_{F}$ is determined by the condition $%
\sum_{n_{x}=0}^{n_{F}}K_{x}=N$, where $N$ is the number of trapped 
atoms. The factor $K_{x}$ reflects the trap geometry. 
For traps with
a cylindric symmetry $\left( \lambda _{y}=\lambda _{z}=\lambda \right) $, one finds $%
K_{x}=\left(\left[ q_x  \right] +1\right) \left(\left[ q_x  \right] + 2\right) /2$, 
where $q_x =\left( n_{F}-n_{x}\right) /\lambda $.
For an anisotropic trap of a cigar shape 
$\left( \lambda_{y},\lambda _{z}\gg 1 \right) $, one has $K_{x}=1$ provided
the number of trapped atoms is small enough $\left(n_{F} < \lambda_{y},\lambda_{z}\right)$. 
In such a situation, the trap becomes
one-dimensional (1D), giving $n_{F}=N-1$. On the other hand, for an
isotropic three-dimensional (3D) trap $\left( \lambda _{y}=\lambda
_{z}=1\right) $, one arrives at $K_{x}=\left( n_{F}-n_{x}+1\right) \left(
n_{F}-n_{x}+2\right) /2$. If the number of atoms is sufficiently large 
($n_{F}>\lambda _{y},\lambda_{z}$),
the anisotropic traps of cigar shape $\left( \lambda
_{y},\lambda _{z}\gg 1\right) $ are no longer one dimensional, since the 
Fermi energy is then greater than the energy of the translational
quanta in the $y$ and $z$ directions. Such a situation corresponds to the
recent experiment by DeMarco and Jin \cite{demarco:99}.

\begin{figure}
\leavevmode
\centering
\epsfxsize=3.00in
\epsfysize=2.50in
\epsffile{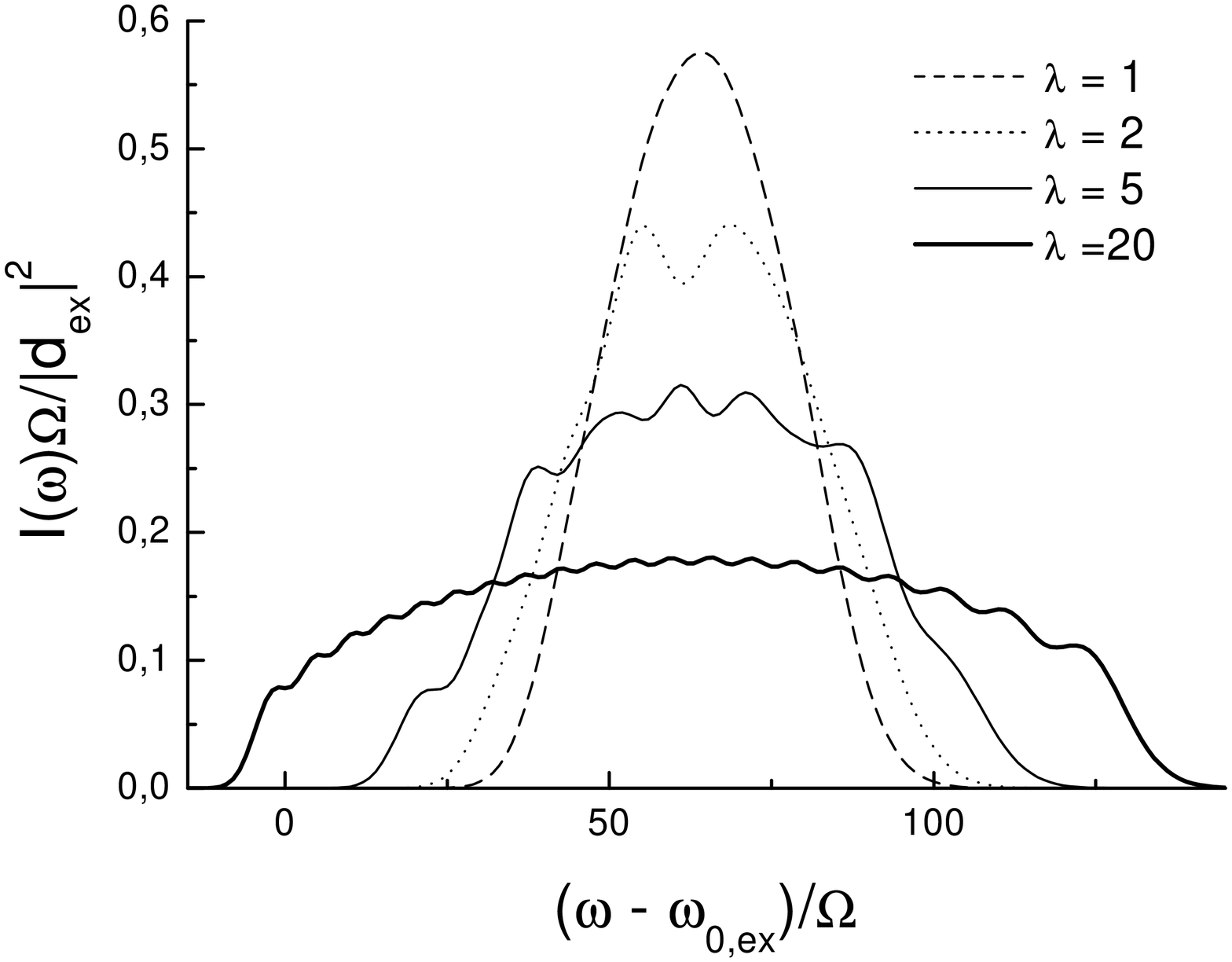}
\caption{Absorption lineshape calculated using Eq.(\ref{I-tiksl}) for $N=20$, $%
T=0$ and $\alpha=8$. The singular contributions due to the delta 
functions have been smoothened.}  
\label{fig:1}
\end{figure}

Figure \ref{fig:1} shows the absorption lineshapes for 
various degrees of the trap anisotropy in the case where 
$\lambda _{y}=\lambda _{z}=\lambda $.
A single excited electronic state $\left| ex\right\rangle $ has been taken
into account in these and the subsequent figures. Oscillations are seen
clearly in the thick solid line representing a
purely 1D case ($\lambda =20$), as well as in the thin one 
corresponding
to an anisotropic 3-D case ($\lambda =5$). This can be
related to the oscillations of the density of Fermi atoms in the
one-dimensional \cite{Gleis:00} and anisotropic three dimensional traps \cite
{Wallis} at a sufficiently small number of the trapped atoms. 
Oscillations do not appear in the lineshape of an isotropic 3D trap (%
$\lambda =1$). Note that in contrast to a single trapped
atom \cite{Gajda:99}, the  zero-temperature lineshape of the trapped 
Fermi gas has a cut off
at the frequencies smaller than $\omega _{0,ex}$. 
This can be explained by the fact that fermions occupy excited translational levels of the trap
 at T=0 (up to the Fermi level), so that optical 
absorption can be accompanied by
a decrease in the translational energy of the atoms.

When $N$ or $T$ is increasing, the behavior of a quantum system becomes more
similar to that of the classical one. To get analytical formulas for the
absorption lineshape at arbitrarily large values of $N$ and $T$, we shall
make use of the semiclassical Thomas-Fermi (TF) approximation. In the TF
approximation, the state of an atom is labeled by the radius-vector ${\bf r}$
and wave vector ${\bf k =\bf p /\hbar }$ (see e.g. refs.\cite{Butts:97,Wallis}). The
density of such states in the six-dimensional phase space $\left( {\bf r},%
{\bf k}\right) $ is $\left( 2\pi \right) ^{-3}$. The number density of the
fermion atoms in the phase space is: 
\begin{equation}
\rho \left( {\bf r},{\bf k},T\right) =\frac{1}{\left( 2\pi \right) ^{3}}\frac{1}{%
\exp \left\{ \beta H_g \left( {\bf r},{\hbar \bf k}\right) -\beta \mu \right\}
+1},  \label{eq:7}
\end{equation}
where the chemical potential $\mu $ is related to the number of trapped atoms via the 
normalization condition 
$\int {\rm d}^{3}{\bf r} {\rm d}^{3}{\bf k}\rho \left( {\bf r},{\bf k}%
,T\right)=N $.

Applying the TF approximation, 
the lineshape reads: 
\begin{equation}
I\left( \omega \right) =\sum_{ex}\left| d_{ex}\right| ^{2}\int%
{\rm d}^{3}{\bf r}{\rm d}^{3}{\bf k}\rho \left({\bf r}, {\bf k},T\right)%
\delta \left( \omega - \omega _{{\bf r}, {\bf k}}%
\right)  \label{eq:8},
\end{equation}
where  $\omega _{{\bf r}, {\bf k}}=%
M \left( \Omega_{ex}^{2} - \Omega _{g}^{2} \right)%
\left( x^{2}+\lambda _{y}^{2}y^{2}+\lambda _{z}^{2}z^{2} \right)/2\hbar%
+\hbar k_x \kappa/M +\hbar \kappa^2 /M + \omega _{0,ex}$
is the transition frequency.
If $\Omega _{g} =\Omega _{ex}$, the frequency 
$\omega _{{\bf r}, {\bf k}}$
does not depend on the atomic position ${\bf r}$, so the lineshape is determined
exclusively by the momentum distribution function: 
\begin{equation}
n\left( {\bf k},T\right) =\int {\rm d}^{3}{\bf r}\rho\left( {\bf r},{\bf k}%
,T\right).  \label{eq:9}
\end{equation}
For $T=0$ the distribution function is given by \cite{Butts:97}: 
\begin{equation}
n\left( {\bf k},0\right) =\frac{8N}{\pi ^{2}K_{F}^{3}}\left( 1-\frac{k^{2}%
}{K_{F}^{2}}\right) ^{3/2},  \label{eq:10}
\end{equation}
where $K_{F}=\left( 2ME_{F}/\hbar ^{2}\right) ^{1/2}$ is the maximum
momentum of the trapped Fermi atoms at zero temperature, 
and $E_{F}=\hbar \Omega
_{g}\left( 6 \lambda _{y}\lambda _{z}N\right) ^{1/3}$ is the
Fermi energy.  
Note that the momentum
distribution is isotropic even though the trap is anisotropic \cite{Butts:97}. 
This leads to an isotropic absorption lineshape in the case where 
$\Omega_{ex}=\Omega_{g}$.
Applying the distribution function (\ref{eq:10}), 
one arrives at the $%
\left(1-z^2\right)^{5/2}$ behavior of the lineshape
if $\Omega_{ex}=\Omega_{g}=\Omega $:
\begin{equation}
I\left( \omega \right) =\frac{16N}{5\pi \Delta }\sum_{ex}\left|
d_{ex}\right| ^{2}\left[ 1-\left( \omega -\omega _{max}\right) ^{2}/\Delta
^{2}\right] ^{5/2},  \label{eq:11}
\end{equation}
where
the central frequency $\omega _{max }=\omega _{0,ex}+\alpha^2 \Omega$ is
shifted by the recoil frequency $\omega_{rec}=\alpha^2\Omega$ as compared 
to $\omega _{0,ex}$, 
\begin{equation}
\Delta =\alpha \Omega \left( 6N\lambda _{y}\lambda _{z}\right) ^{1/6}
\label{eq:12}
\end{equation}
being the spectral half-width. In the experiment by DeMarco and Jin 
\cite{demarco:99} using trapped $^{40}$K atoms, 
$\Omega =2\pi \times 19$ Hz, $N=7\times 10^{5}$ and $%
\lambda \approx 7$, giving $\alpha \approx 36$ for $\omega=4\times 10^{16}$ Hz. 
Consequently one has
$\Delta \approx 6\times 10^{5}$ Hz. This is less
than the typical radiative linewidths for free atoms. Yet, for trapped
fermions the spontaneous emission is suppressed \cite{Busch}, so
the Doppler broadening can be dominant.

The approximated lineshape (\ref{eq:11}) 
depends on the trap anisotropy exclusively
through the characteristic frequency 
$\Omega _{char}=\Omega \left( \lambda _{y}\lambda _{z}\right) ^{1/3}$ 
which is a measure of
the trap hardness. The bigger $\Omega _{char}$ is, the tighter is the trap,
and the broader is the absorption spectrum. In fact, the maximum momentum of
the atoms is larger in tighter traps (for the same number of trapped atoms)
leading to the increase in the Doppler broadening. For instance, compared to
an isotropic trap $\left( \lambda _{y}=\lambda _{z}=1\right) $, the spectrum
of a squeezed trap ($\lambda _{y}$, $\lambda _{z}>1$) is broader.

\begin{figure}[t]
\leavevmode
\centering
\epsfxsize=3.00in
\epsfysize=2.50in
\epsffile{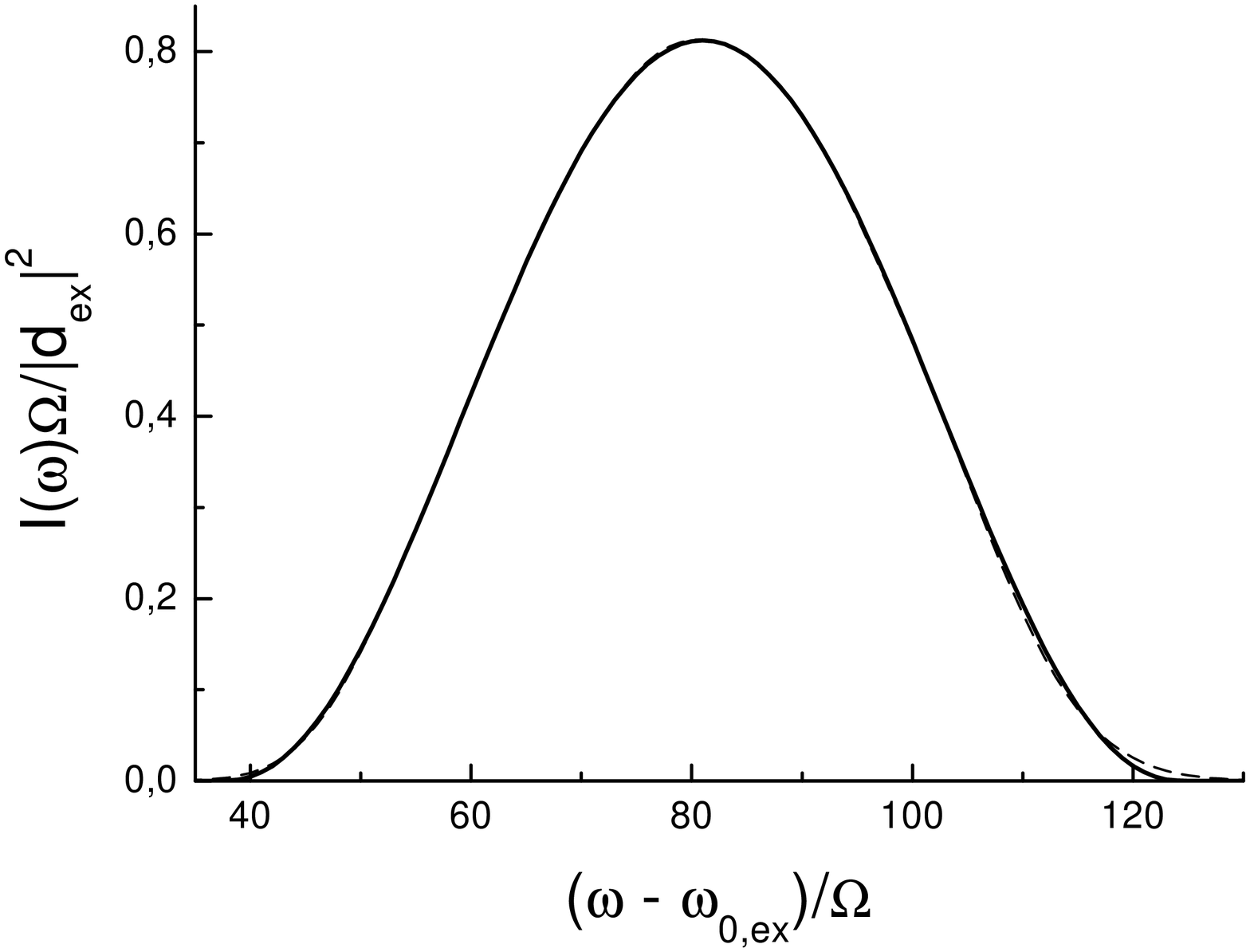}
\caption{Absorption lineshape at $T=0$ calculated using the exact summation
(dashed line) and the TF approximation (solid line) for $\alpha=9$, 
$N=35$ and $\lambda_x=\lambda_y=\lambda =1$.}
\label{fig:2}
\end{figure}

\begin{figure}
\leavevmode
\centering
\epsfxsize=3.00in
\epsfysize=2.50in
\epsffile{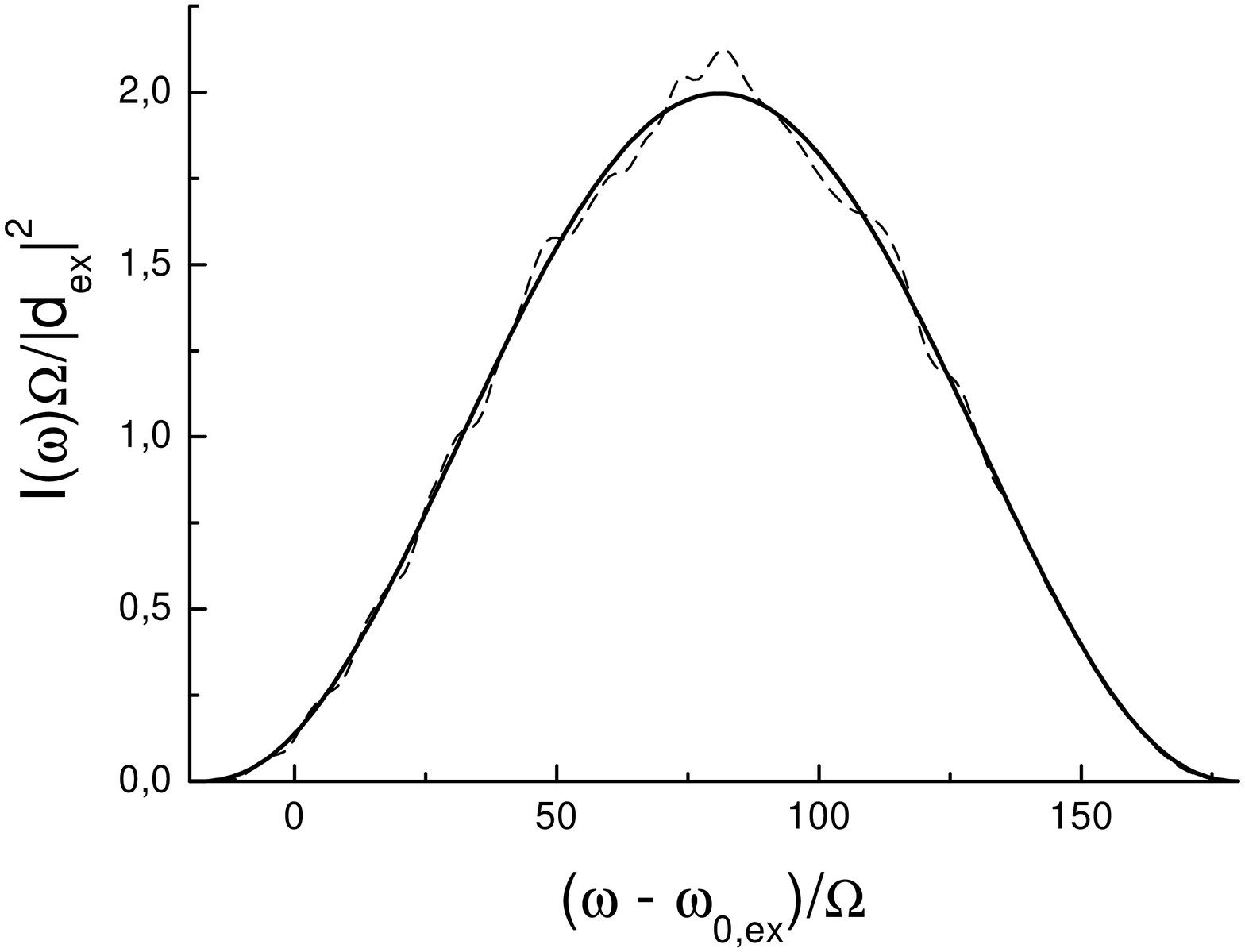}
\caption{The same as in Fig. \ref{fig:2}, but for $N=196$ and $\lambda=5$.}
\label{fig:3}
\end{figure}

Exact and approximated lineshapes are presented in Figs. \ref{fig:2} and \ref{fig:3}. For an
isotropic trap (Fig. \ref{fig:2}), the agreement appears to be very good, even though the
number of atoms $N$ is rather small. Deviations are seen only in the tails of the spectrum
corresponding to the periphery of the fermion cloud. In such an area, the fermion 
density becomes small and  the TF approximation fails \cite{Butts:97}. For 
anisotropic 
traps, the exact spectrum undergoes some oscillations about the approximated 
one even for relatively large
values of N (see Fig. \ref{fig:3}). 
In fact, the energy of translational quanta depends now on the specific
directions of atomic motion, so a larger number of trapped atoms is needed
to populate substantially the translational levels in all three directions.

Consider next the situation where $T\neq 0$ and the frequencies $\Omega _{g}$ and 
$\Omega _{ex}$ are not necessarily equal. The lineshape (\ref
{eq:8}) takes then the form for an isotropic trap: 
\begin{eqnarray}
I\left( \omega ,T\right) &=&\sum_{ex}\frac{\left|
d_{ex}\right| ^{2}}{16\pi\alpha^6\Omega^6 p^{5/2}}\int_{0}^{\infty }y^{2}{\rm d}y 
\ln \left\{ 1+\exp \left[\beta \mu   \right.\right.  \nonumber \\
&&\left. \left. -y^{2}-p\left(
\omega -\omega _{max} + m_{ex}y^2/\beta\hbar\right)^{2}\right] \right\}, 
\label{eq:13}
\end{eqnarray}
where $m_{ex}=1-\Omega _{ex}^{2}/\Omega _{g}^{2}$, \quad 
$p=\beta \hbar /4\alpha ^2\Omega $ and $\Omega\equiv \Omega _{g}$.
If $\Omega_{g} = \Omega _{ex}$, the result (\ref{eq:13}) 
can be extended readily to anisotropic traps. 
In such a case, the lineshape (\ref{eq:13}) acquires an extra factor 
$1/ \lambda_x \lambda_y $, and the chemical potential $\mu $  
depends on $\lambda_x \lambda_y $, in addition to $T$ and $N$.    

We are interested primarily in the strongly degenerate Fermi
gas $\left( {\beta }{E_{F}}\gg 1\right) $, for which the Sommerfeld expansion
holds for the chemical potential:
\begin{equation}
\mu =E_{F}\left( 1-\frac{\pi ^{2}}{3}\left( \frac{1}{\beta E_{F}}\right)
^{2}\right)  \label{eq:14}
\end{equation}
In the opposite (non-degenerate gas) limit 
$\left( {\beta }{E_{F}}\longrightarrow 0\right) $, 
one has $\mu =\beta ^{-1} \ln \left[
\left( \beta E_F\right) ^{3}/6\right] $, and
the lineshape (\ref{eq:13}) reduces to the Gaussian form
if $\Omega_{g} = \Omega _{ex}$. 

\begin{figure}
\leavevmode
\centering
\epsfxsize=3.00in
\epsfysize=2.50in
\epsffile{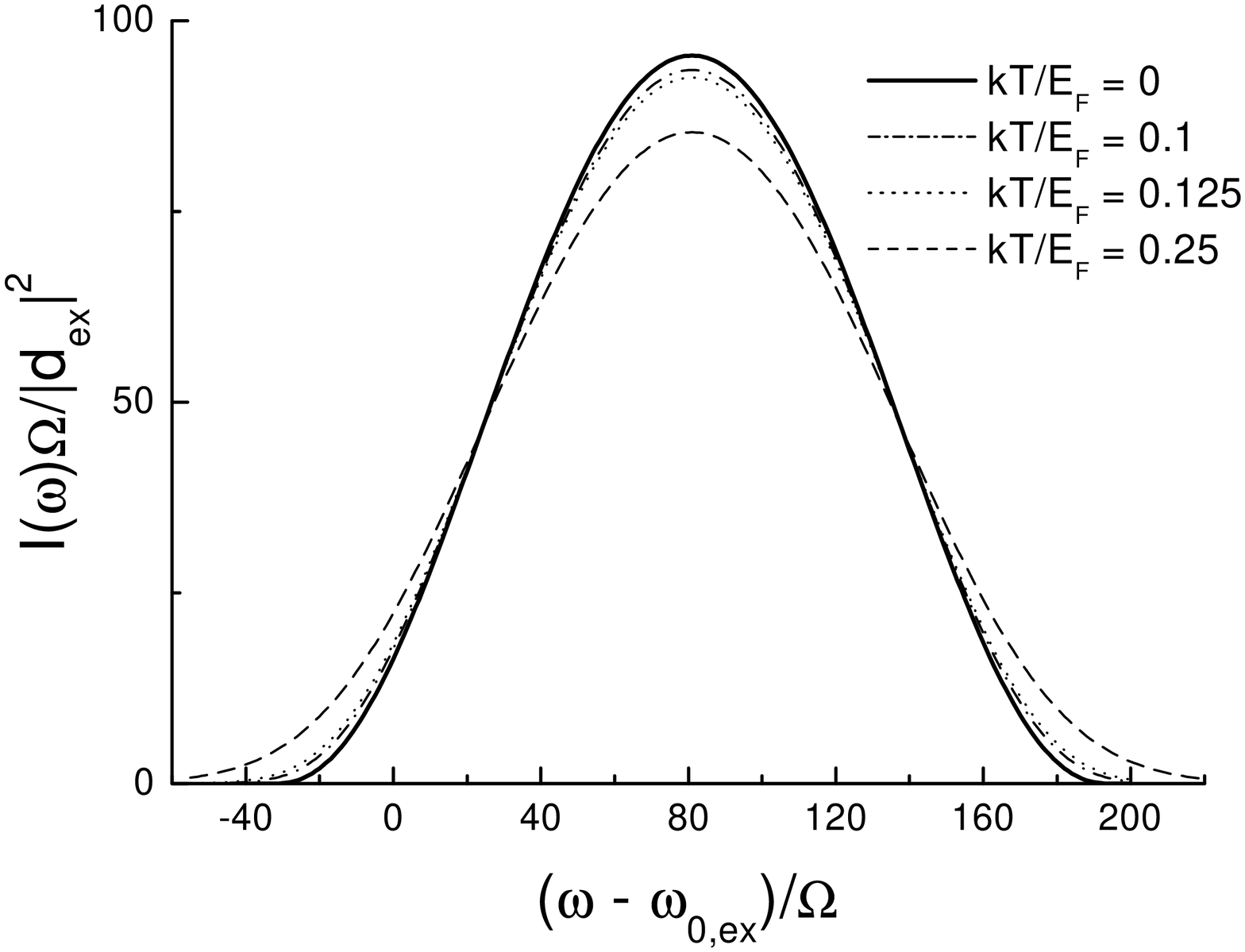}
\caption{Absorption lineshape at
various temperatures for an isotropic trap with 
$\Omega _{g}=\Omega _{ex}$, $N=10667$ and $\alpha=9$.}
\label{fig:4}
\end{figure}

\begin{figure}
\leavevmode
\centering
\epsfxsize=3.00in
\epsfysize=2.50in
\epsffile{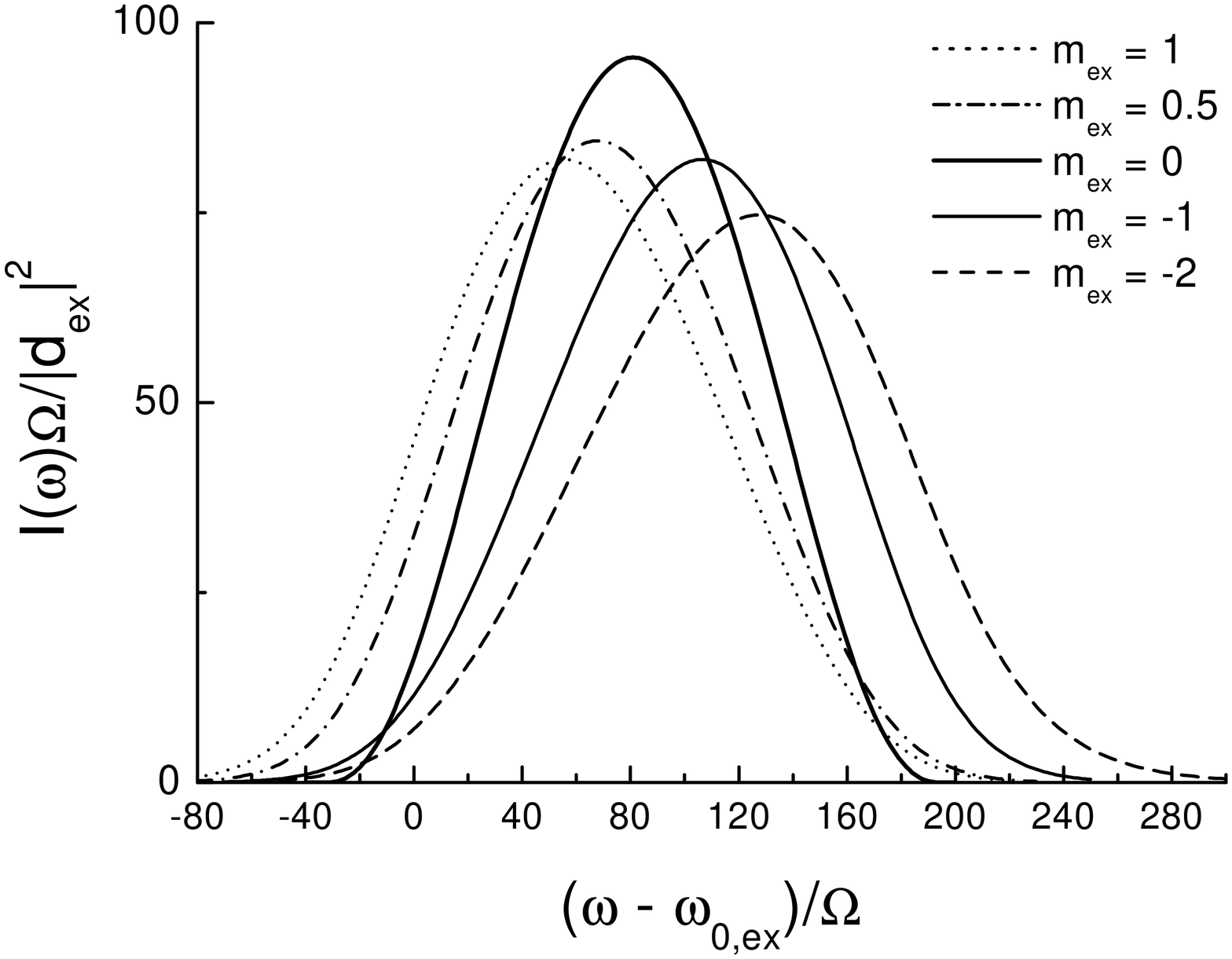}
\caption{Absorption lineshape for an isotropic trap with $N=10667$, 
$\alpha=9$, $k_{B}T=0.25E_{F}$ and various
$\Omega _{ex}/\Omega _{g}$.}
\label{fig:5}
\end{figure}

Figure \ref{fig:4} shows the temperature dependence of the lineshapes 
for $\Omega _{g}=\Omega _{ex}$. 
At very low temperature ($kT/E_{F}=0.1$), the
absorption spectrum is seen to be close to the zero-temperature limit. As
the temperature increases, the absorption lineshape becomes broader and is
no longer characterized by the $\left( 1-z^{2}\right) ^{5/2}$
behavior. Yet, the lineshape is still non-Gaussian, since the gas is
strongly degenerate. Figure
\ref{fig:5} shows the absorption spectrum for various values of $m_{ex}=1-\Omega
_{ex}^{2}/\Omega _{g}^{2}$. For $\Omega _{ex}>\Omega _{g}$, the maximum
position of the spectrum is shifted to larger frequencies. For $\Omega
_{ex}<\Omega _{g}$, one has the opposite. Furthermore, one can see the
obvious increase in the spectrum width if $\Omega _{ex}>\Omega _{g}$. This
is due to the fact that an increase in the translational frequency of the
electronically excited atoms leads to an increase and broader distribution
of frequencies of the optical transitions.

In summary, we have studied the absorption spectrum by a cold gas of Fermi
atoms using both the exact summation and also the Thomas - Fermi
approximation. Oscillations have been obtained in the absorption spectrum
calculated numerically for one-dimensional and anisotropic three-dimensional
traps at a sufficiently small number of trapped atoms and $T=0$. No such
oscillations appear for the isotropic three-dimensional traps. Applying the
TF approximation, relatively simple analytical expressions have been
obtained for the lineshape of three dimensional
traps at a sufficiently large number of trapped particles. At zero temperature, 
the approximated spectrum is characterized
by a $\left( 1-z^{2}\right) ^{5/2}$ dependence. 
At non-zero temperature, the spectrum becomes broader, although
remains non-Gaussian as long as the fermion gas is degenerate. The changes in the 
trap frequency for an electronically excited atom can introduce 
an additional line broadening.

The authors would like to express their appreciation for the useful discussions 
with H. Carmichael and T. Mossberg. This work has been completed during 
the Fulbright Scholarship by G. J. at the University of Oregon.

\end{document}